\documentclass[12pt]{article}

\usepackage{newtxtext,newtxmath}

\usepackage{graphicx}

\usepackage[letterpaper,margin=1in]{geometry}

\renewenvironment{abstract}
	{\quotation}
	{\endquotation}

\date{}

\makeatletter
\renewcommand{\fnum@figure}{\textbf{Figure \thefigure}}
\renewcommand{\fnum@table}{\textbf{Table \thetable}}
\makeatother

\usepackage{url}

\def\mytitle{
Code evolution for link prediction in complex networks
}

\title{\bfseries \boldmath \mytitle}

\author{
	Alexey Vlaskin$^{1\ast}$,
	Eduardo G. Altmann$^{1}$\and
	\small$^{1}$ School of Mathematics and Statistics \& Centre for Complex Systems\and
    \small The University of Sydney, 2026, NSW \& Sydney, Australia.\and
	\small$^\ast$Corresponding author. Email: alex@avlaskin.com
}

\begin{document}

\maketitle

\begin{abstract} \bfseries \boldmath
The problem of predicting links in complex networks appears in different disciplines and has led to a variety of ingenious human-designed methods. We use this rich program space to explore the performance and behaviour of automated code-evolution systems tasked to obtain machine-designed methods for link prediction. Despite being trained on limited data, algorithms evolved through
code evolution outperform human-designed methods (with an average AUC score of 0.915 vs. 0.783, computed over 580 networks) and show improved computational efficiency, allowing them to be applied to networks with millions of links. The discovered methods follow approaches that have been employed in human-designed methods, but contain key innovations in the selection and combination of node- and link-features. This illustrates the role modern large language models and genetic algorithms can play in algorithmic innovation and scientific discovery more generally.

\end{abstract}

\noindent

\section{Introduction}

The combination of code-evolution systems~\cite{geneticompute} and large language models (LLMs) ~\cite{brown2020language} has opened new promising approaches for automating the discovery of scientific algorithms~\cite{funsearch,alphaevolve,shinkaevolve,codeevolve}. Traditionally, developing complex scientific algorithms required human experts to iterate, evaluate, and refine algorithms. In code-evolution systems, this iterative process is  automated and has led to successful applications to a variety of mathematical~\cite{mathematical,mathresearch} and NP-hard~\cite{nphard} problems. The mixed success reported in Ref.~\cite{mathematical} shows that the general applicability and potential of this approach is unclear.
Exploring the application of these automated code-evolution approaches to new, complex domains is thus essential to understand their capabilities, their limitations, and the extent to which their generated solutions differ from traditional human-designed methods.

Here we assess the applicability of automatic algorithm discovery to the problem of link prediction in complex networks. Developed over the last several decades, link prediction aims to use existing network information to predict unseen or missing links~\cite{prauc}.
The topic has been explored by diverse scientific communities — ranging from network science and machine learning to biology and sociology—yielding a vast and varied landscape of algorithmic solutions, ranging from similarity measures such as the Jaccard index ~\cite{jaccard} and Adamic-Adar index ~\cite{adamic}, to graph neural networks (GNNs)~\cite{GNNModel, GraphSage, SEAL}, random-walk methods such as Node2Vec~\cite{node2vec}, structural similarity metrics~\cite{DLS2020}, and probabilistic generative models such as Stochastic Block Models (SBM)~\cite{SBM,survey}. These diverse methodologies, frequently tailored to their respective fields -- such as social, transportation, economic or biological networks—create an exceptionally rich program space. Consequently, link prediction serves as a uniquely suited environment for evaluating code-evolution systems. Here we investigate the capabilities of code evolution through a two-fold lens: exploring the evolution of algorithms and conducting an analysis of the resulting discovered methods to see how they consolidate or improve upon human-designed methods.

The importance of link prediction in network science is that it uncovers hidden relationships and forecasts the future evolution of complex systems.
It provides a quantitative test of the ability of algorithms to describe and exploit regularities in the network structure. In applied settings, identifying likely connections that have not yet formed or are simply unobserved can improve recommendation engines~\cite{lu2011link}, optimise biological protein-protein interaction maps~\cite{kovacs2019network}, and even assist in criminal investigations by finding missing links in criminal networks~\cite{berlusconi2016link}.
The core difficulty in link prediction lies in designing algorithms capable of exploring diverse statistical regularities—such as degree variability, motifs, and meso-scale structures—at different network scales.
These and other regularities are observed in complex networks across different domains~\cite{barabasi_network_2016}, suggesting that a single well-designed link-prediction algorithm can be successful in a wide range of networks. At the same time, it is important to recognise the intrinsic unpredictability of complex networks~\cite{jing2026predictability} and the existence of no-free lunch theorems in related problems in complex networks~\cite{nflt,lu2011link,peel2017ground}, which indicate that no single algorithm can be optimal in all networks.
Therefore, instead of pursuing the elusive goal of developing a universal optimiser, we aim at generic link-prediction methods that succeed in a variety of (large, sparse) complex networks.

In this paper, we propose a methodology that uses a code-evolution system, guided by LLMs, to automatically discover new link-prediction methods. To train and evaluate the evolution system without overfitting to specific domains, we use a small set of networks which combine empirical and synthetic networks~\cite{vlaskin2025synthetic}. We then rigorously assess the machine-discovered algorithms on previously-proposed test sets, which include 580 real networks~\cite{pnas2020,eval2019} from various fields. Our goal is not only to obtain improved link-prediction methods but also to answer three primary research questions:
\begin{enumerate}
  \item To what extent can code evolved algorithms generalise across diverse topologies without becoming overly specialised? To answer this question, we compare the performance of the best algorithms on diverse datasets to different human-designed methods.
  \item How critical is the evolutionary process to obtain high-performance link-prediction methods? We address this via ablation studies focused on code evolution.
  \item Does the evolutionary process produce genuinely novel algorithms or does it merely consolidate existing human knowledge? We analyse hundreds of top-performing algorithms to answer this question.

\end{enumerate}

\section{Methods and materials}

\subsection{The Link prediction problem}

We study networks made up of nodes connected by (directed) links, which are divided into two disjoint sets: observed links that we know exist and unobserved links that are currently hidden from us. The link-prediction problem we consider here consists in identifying the unobserved links using the observed links.  For simplicity, we assume all nodes of the network appear in the observed links and are thus known.

Link-prediction methods assign a probability or rank to all links in the set of potential links, which consist of the (typically smaller) set of unobserved links and the set of non-existing links.  In most network datasets, the distinction between observed and unobserved links is not present.  Therefore, in practice, the evaluation of link-prediction methods is performed by randomly dividing all links into the two disjoint sets of observed and unobserved links.
In line with most previous works~\cite{graphevalnn, vlaskin2025synthetic, pnas2020, DLS2020},
here we choose $10\%$ to $20\%$ of the links of a network to be unobserved and we quantify the quality of link-prediction methods using the Area-Under-the-Curve (AUC) score. AUC measures the overall probability that the model will correctly rank a randomly chosen positive link higher than a randomly chosen negative one and it evaluates a model's performance across all possible classification thresholds simultaneously. So AUC is threshold free, normalised $AUC \in [0; 1]$, and it has a well-defined value $AUC=0.5$ for a random prediction (null model). The AUC score of a method in a network is computed as the average over multiple realisations of the random deletions, and all methods are evaluated on the same set of links, see Supplementary Materials (SM) Sec.~\ref{app:auc} for details on the methods evaluation.

\subsection{Code-evolution system}\label{sec:codevolve}

Here we describe our code-evolution system to generate link-prediction algorithms, which uses the AlphaEvolve system proposed in Ref.\cite{alphaevolve} as a blueprint. The smaller size of our system motivates us to call it AntEvolve or AE.

AntEvolve's process of code evolution starts with an initial program and a human-designed evaluator. The initial program $p_0$ defines the necessary interface for the evaluation protocol without providing the functional logic of link prediction. As evaluator score (fitness), we use the AUC score of a program $AUC(p)$ obtained on ten unique networks: six synthetic networks from Ref.~\cite{vlaskin2025synthetic} and four real networks from Ref.~\cite{tiago2020} (see SM Sec.~\ref{app:training} for details). The initial program ranks potential links randomly and thus $AUC(p_0)=0.5$.

At some evolutionary time $t$, new programs $\bar{p}$ are proposed based on a subset $S_t$ of all previously accepted programs $A_t$, i.e., $S_t \subseteq A_t$. Any program $p'$ proposed at a time $t'<t$ can be included in $S_t$, but preference is given to programs $p$ with high fitness $AUC(p)$. Our default choice is to have $S_t$ composed of the ten $p' \in A_t$ with highest $AUC(p')$.
Two strategies are used to preserve genetic diversity and mitigate premature convergence (monopoly): (i) with a probability of 1\% a random program $p' \in A_t$ is selected as the sole program in $S_t$;
(ii) for $t<K=200$, the global population is partitioned into $10$ discrete sub-populations, or islands, following Ref.~\cite{tanese1989distributed}, and the ten programs included in $S_t$ are restricted to be from the same island. The code evolution process can be slow as the system spends most of the time evaluating programs. To speed up the process, we use $F=10$ parallel flows that work simultaneously,
accessing and contributing to the same pool of accepted programs $A_t$ (for $t < K=200$, each flow works exclusively on one assigned island).

We now describe the generation of new programs $\bar{p}$ from the set of selected programs $S_t$. The key step is to task an LLM to propose a program that increases $AUC(p)$ based on changes to the programs in $S_t$. Two approaches (selected with equal probability) are used: {\it mutation} of a single randomly selected program from $S_t$;  and {\it crossover} of three randomly selected programs from $S_t$. This shortlist lottery is designed to balance meritocracy (selecting high performers) with fairness and diversity (reducing bias and winner-take-all outcomes). The described code evolution process allows for sequential program improvement because $S_t$ will typically have the top performing programs found so far.
The specific LLM used for mutation/crossover is sampled from a set (e.g. the Google Gemini ~\cite{gemini25} Flash 3 preview with 90\% probability and the Gemini 3 Pro preview with 10\% probability).
In order to be accepted into the set of programs $A$, the program $\bar{p}$ -- produced by mutation or crossover -- must satisfy a structural validity constraint, performed as a code static analysis via \texttt{pylint}. Programs are accepted if they are new, free of linting errors, and are evaluated within a maximum evaluation time ($<1200$ seconds). If the program $\bar{p}$ is rejected due to errors, taking too long to evaluate, or being a duplicate of a program in $A_t$, it is discarded, and the process starts again. Accepted programs are added to $A$ (irrespective of their AUC score) and can thus be selected for further evolution.
The programs  $p_i$, with $i=0,1,2,\ldots, |A|$, are indexed according to the time they were added to $A$ and the evolution process stops when a pre-defined number $M$ of programs are accepted ($|A_t| \ge M$). The best performing algorithm is chosen as
$$p^* = p_{i^*} \quad \text{where} \quad i^* = \underset{i \in [0,M]}{\operatorname{argmax}} \text{AUC}(p_i).$$
\begin{figure}[!bt]
	\centering
    \includegraphics[width=0.7\textwidth]{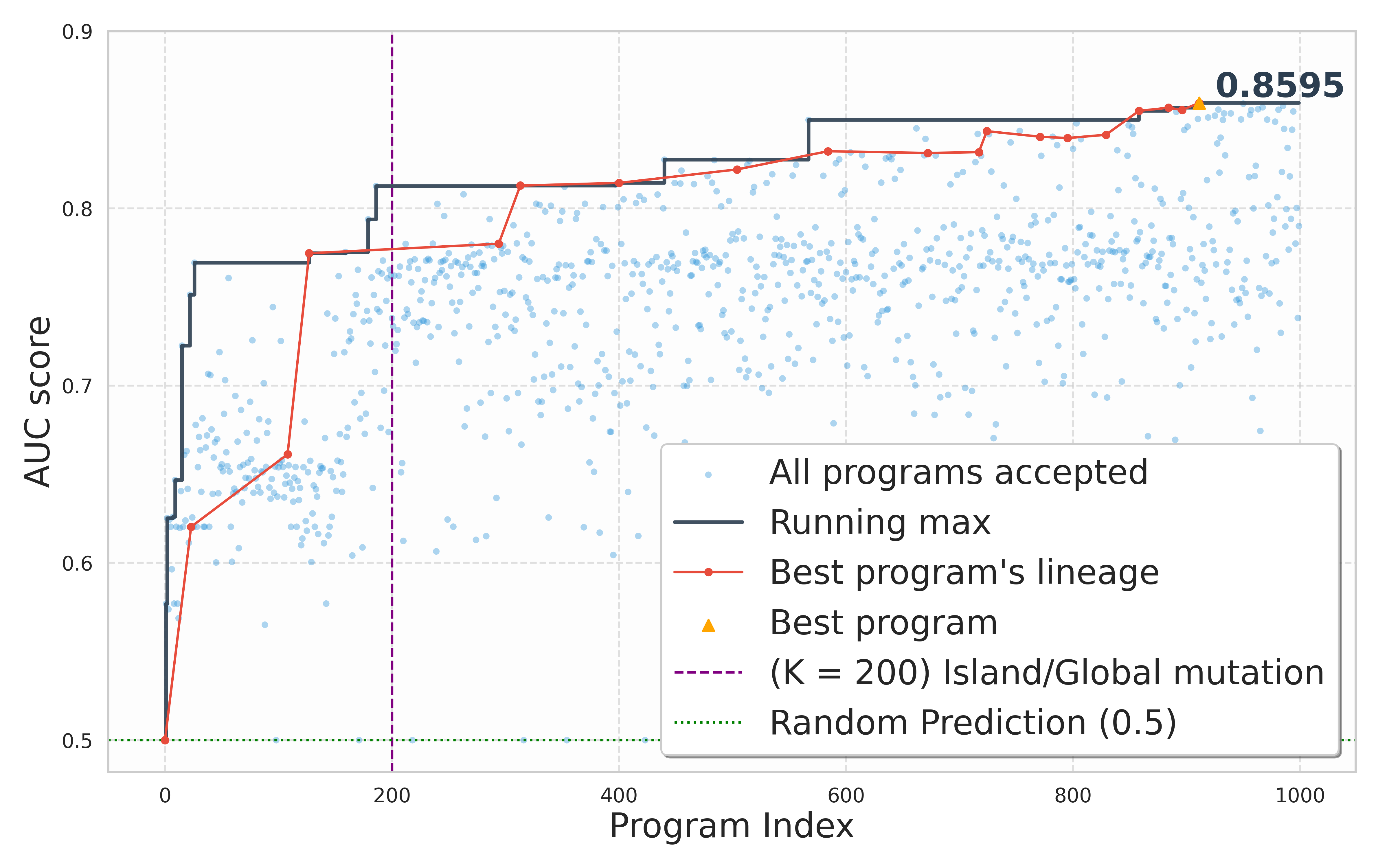}
	\caption{\textbf{Code evolution process.}
	  Fitness (AUC score, y-axis) of the programs $p_i$ found at different time steps (program indexes $i$, x-axis) during the code evolution.
     The highest-score program $p^*$ (orange symbol) and its ancestry lineage (red line) are shown. The ancestry lineage connects a program to the top performing parent program in the previous generation. }
	\label{fig:code_evo}
\end{figure}

 \begin{table}[!htbp]
\centering
\begin{tabular}{|c|c|c|}
\hline
\textbf{ \/ } & \textbf{Gemini } & \textbf{Qwen}  \\ \hline
Accepted programs & 1002 &  1002 \\
Completion Tokens & 9,754,147 & 33,037,448  \\
Generations & 1221 & 2748  \\
Mutation success rate & 82\% &  36\% \\
Experiment Cost & \$52.50 &  \$9.60 \\
\hline
Training AUC score of best program $p^*$ & 0.896 & 0.878  \\
Number of features used in $p^*$ & 24 & 40 \\ \hline
\end{tabular}
\caption{\textbf{Gemini vs Qwen code evolution comparison.} Comparison of two code evolution runs, both using the same training networks and stopping criterion ($M=1,002$) but different LLM. }
\label{tab:models_table}
\end{table}

Figure~\ref{fig:code_evo} shows one realisation of the process of code evolution obtained using Google Gemini~\cite{gemini25} as the LLM. We see how the fitness of programs $AUC(p_t)$ tends to increase over time $t$, with increasingly rare record-breaking events. Interestingly, low-scoring programs continue to be generated even if significantly better programs exist, also after the sub-population partition stops ($t>K$), suggesting that the system continues to explore wide regions of the program space. Similar results were obtained using Qwen3~\cite{qwen3coder} as an open-source alternative LLM (Qwen3-Coder-30B-A3B is sampled with 90\% probability and Qwen3-Coder-Next with 10\% probability). Table~\ref{tab:models_table} compares these two runs of AntEvolve in terms of the resources they needed and their success in producing acceptable programs. These results show that, while Qwen is cheaper than Gemini, it required around $3$ times more tokens and generations to get to the same number of accepted programs, leading to longer runs and slightly lower AUC scores achieved during the evolution.
In the next section, we evaluate the two best program $p^*$ -- one for each choice of LLM -- by comparing them with human-designed programs in three ensembles of networks (different from the $10$ used during evolution).

\section{Results}
In order to test the performance of the best programs $p^*$ obtained from AntEvolve,
we compute their AUC score on three ensembles of networks: (1) 8 synthetic networks introduced in Ref.~\cite{vlaskin2025synthetic}, (2) 550 empirical small/medium networks used in Ref.~\cite{pnas2020}, and (3) 30 large real networks extracted from Ref.~\cite{tiago2020} (see SM-Table.~\ref{tab:stat30nets} for the exact description of all 30 networks). We compare the AUC score and the speed of algorithm evaluation of AntEvolve in these ensembles to the results obtained by four popular human-designed methods: Stacked model, a stacked model that uses various topological features~\cite{pnas2020}; SBM, a degree corrected Stochastic Block Model~\cite{SBM}; Node2Vec, a random walk based algorithm~\cite{node2vec}; and Adamic-Adar, an heuristic approach based on a similarity index~\cite{adamic}. These methods cover key methods used in the modern scientific literature, with Adamic-Adar representing the expected weakest link-prediction method and the Stacked model representing the strongest. To represent the Stacked model method we used the all topological features model because it is easily replicable and one of the best-performing programs proposed~\cite{pnas2020}, which tested about 200 different methods.

\subsection{Synthetic networks}\label{ssec.synthetic}

\begin{figure*}[hbtp]
    \centering
        \begin{minipage}[c]{0.28\textwidth}
        \centering
        \includegraphics[width=\textwidth]{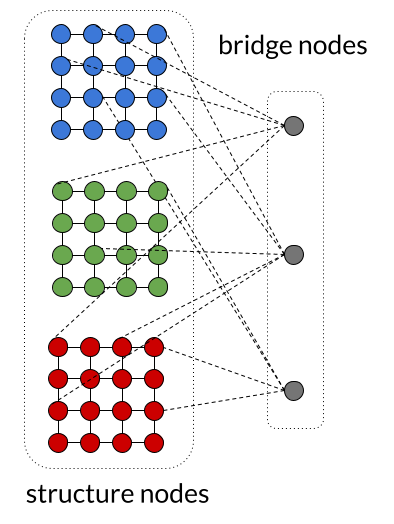}
    \end{minipage}
    \begin{minipage}[c]{0.6\textwidth}
        \centering
        \includegraphics[width=\textwidth]{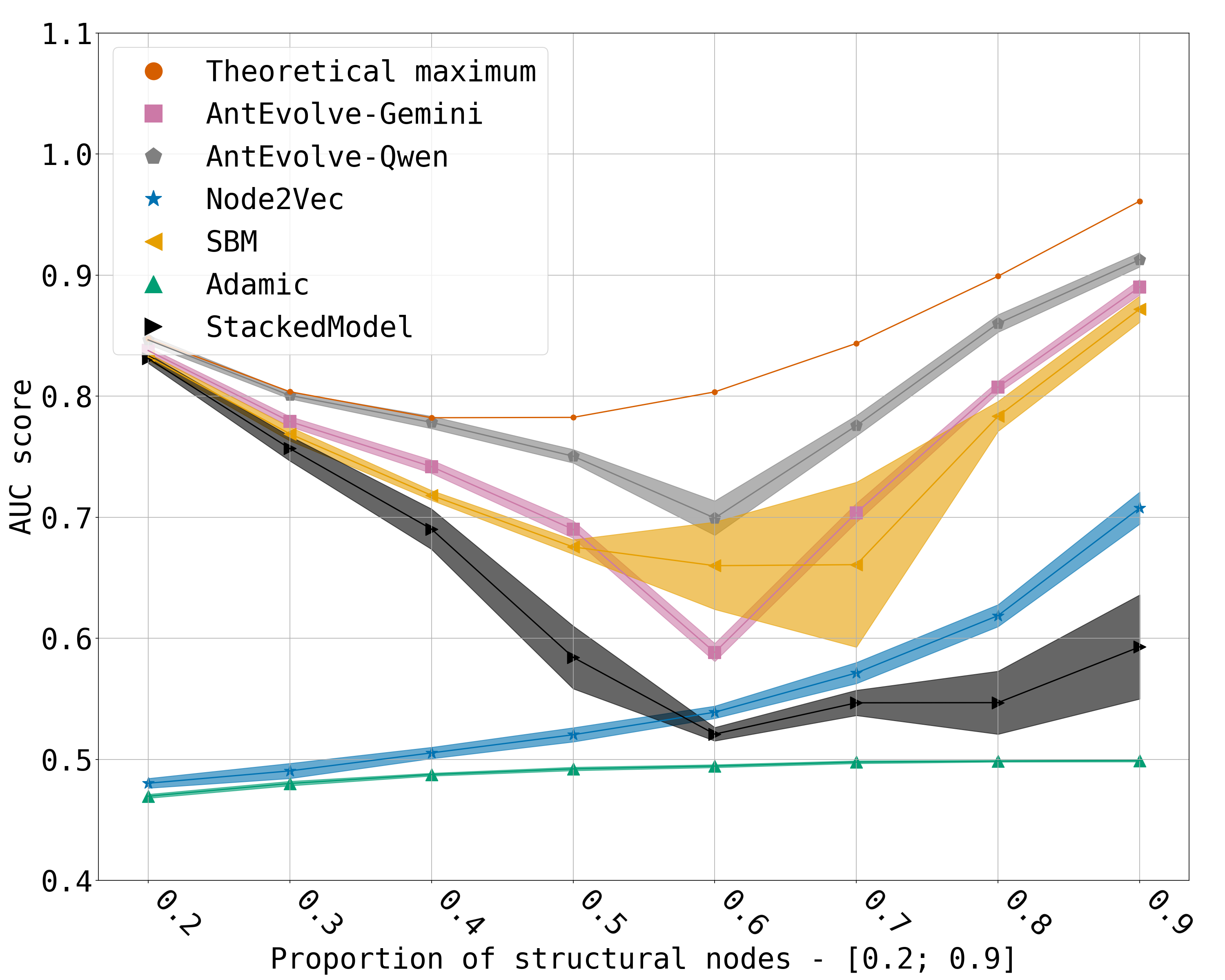}
    \end{minipage}

    \caption{\textbf{Comparison of human-designed and evolved link-prediction methods in synthetic networks.} Left panel: Example of a synthetic network used for evaluation, showing two types of nodes: structure nodes that link deterministically to other nodes in the same structure (a lattice, in this case); and bridge nodes that link to randomly selected structural nodes, see Ref.~\cite{vlaskin2025synthetic} for details. Right panel: Comparison of the performance of six link-prediction methods to the theoretical maximum on synthetic networks with $8$ different network configurations (x-axis). AUC results are computed across 10 instances, with the average and variance shown in the plot. The horizontal axis indicates the proportion of structural nodes among all nodes in the network configuration. All networks have 3,200 nodes, average bridge-nodes degree $12$, and $8 \times 8$-node lattices as structure, with the number of structures increasing from 10 to 45 as the proportion of structural links increases from 20\% (left) to 90\% (right). }
    \label{fig.syntheticNets}
\end{figure*}

Here we use synthetic networks proposed in Ref.~\cite{vlaskin2025synthetic} to evaluate link-prediction methods. Following work in~\cite{vlaskin2025synthetic} we remove 10\% of links from each graph ten times randomly and have the same number of non-existing links as negative examples for link prediction.  This ensemble is designed to capture essential meso-scale and micro-scale network properties in a controlled way, with a computable theoretical maximum AUC score that acts as an upper bound for the performance of any algorithm.
While three networks from this synthetic random-graph model were used during code evolution, here we use ten realisations and different parameters spanning 8 different configurations. The results obtained by the $6$ different link-prediction algorithms are shown in Fig.\ref{fig.syntheticNets}. They show that the best AntEvolve methods $p^*$ outperform the human-designed methods on most configurations and yields results with the same non-monotonic behaviour as the theoretical maximum. The $p^*$ obtained using Qwen achieves scores very close to the theoretical maximum in all configurations.

\subsection{550 real networks}\label{ssec.550}

\begin{figure*}[hbtp]
    \centering
    \begin{minipage}[t]{0.23\textwidth}
        \vspace{0pt}
        \centering
        \includegraphics[width=\textwidth]{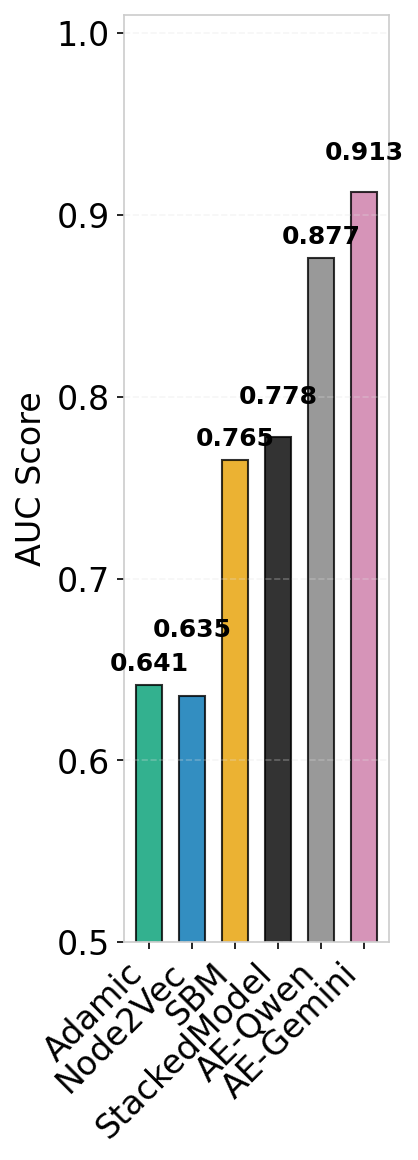}
    \end{minipage}
    \begin{minipage}[t]{0.68\textwidth}
        \vspace{0pt}
        \centering
        \includegraphics[width=\textwidth]{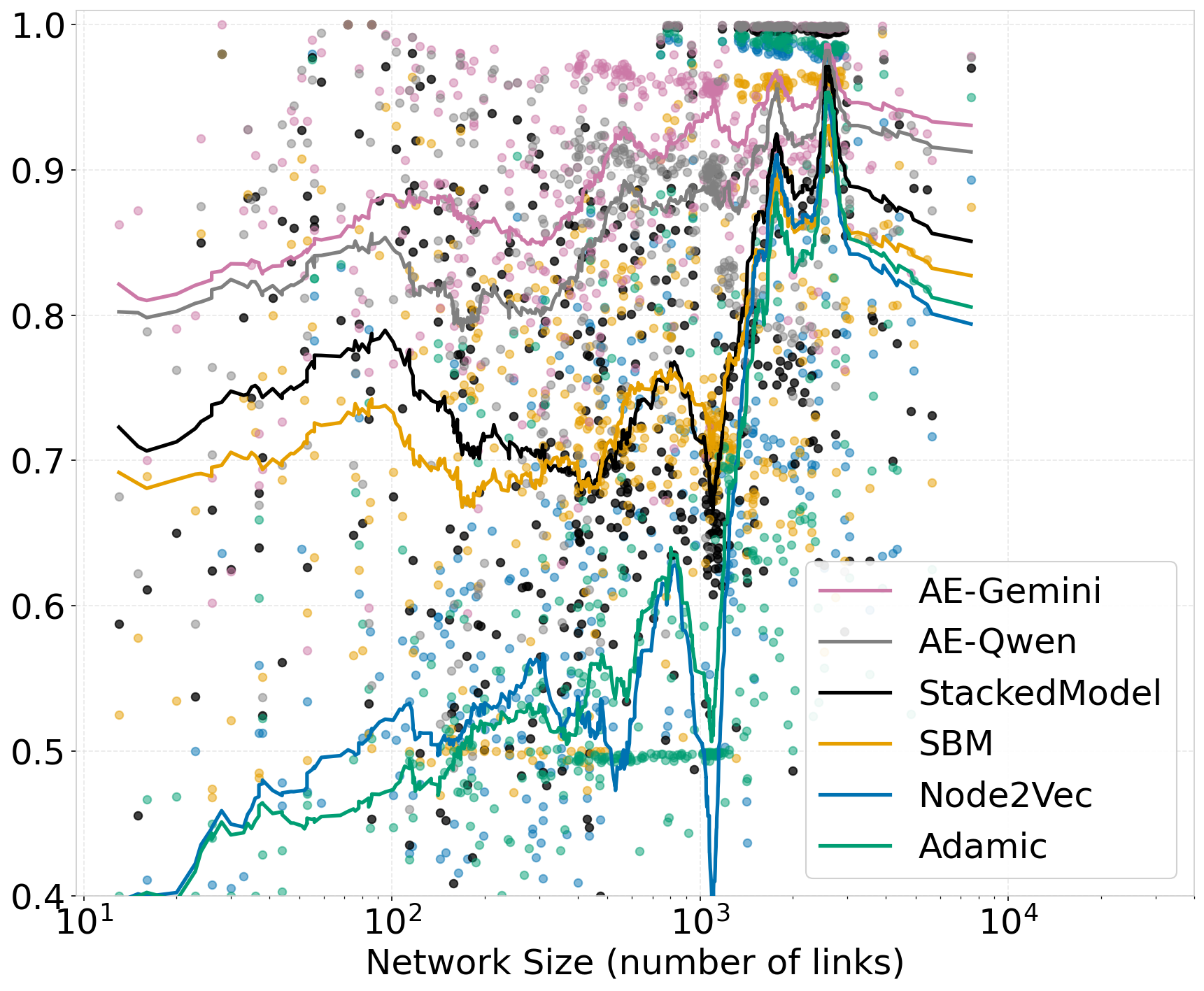}
    \end{minipage}

    \caption{\textbf{Comparison of human-designed and evolved link-prediction algorithms on 550 networks.} Left panel: Overall mean AUC score averaged across $5,500$ graph instances per method. Right panel: Detailed AUC score of methods across $550$ networks of different sizes, ordered by the number of links.
    Each dot represents the average AUC obtained over 10 random samples of links (20\% of the links considered to be unobserved). Lines represent moving average.}
    \label{fig.graphs550}
\end{figure*}

We now evaluate our two machine-evolved programs $p^*$ on empirical networks.
In order to avoid biases in the selection of the networks, we follow Ref.~\cite{pnas2020} and use their entire proposed dataset (available in Ref.~\cite{eval2019}), which has 550 real networks from various domains: biological, social, technological, economic, informational, and transportation networks. We also used similar running average curves to visualise the result shown in Fig.\ref{fig.graphs550}. We sample unobserved links 10 times from each of the 550 networks and present each method with the same set of network instances where $20\%$ of links are missing and none of these networks are seen at training.

There we observe that the best algorithm found by code evolution surpasses all human-developed methods. This includes the  Stacked model (all topological features model) introduced in Ref.~\cite{pnas2020}. While in larger networks the link-prediction problem is typically easier, and high AUC scores are obtained in almost all methods, on smaller networks it becomes challenging, and the higher performance of AntEvolve becomes more evident. In some of these networks, the best machine-designed methods are better than the best human-designed by a large margin, up to 20\% on average.

\subsection{30 large networks}

\begin{figure}
	\centering
    \begin{minipage}[t]{0.23\textwidth}
        \vspace{0pt}
        \centering
        \includegraphics[width=\textwidth]{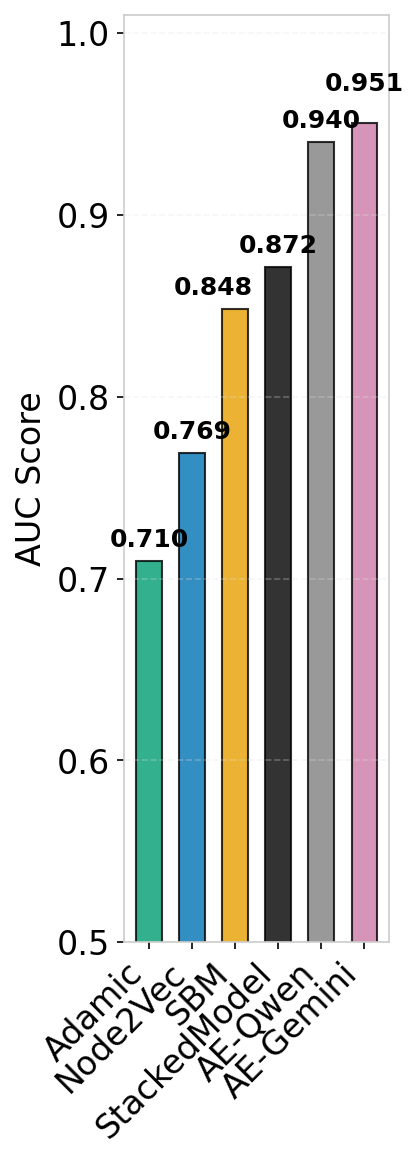}
    \end{minipage}
    \begin{minipage}[t]{0.68\textwidth}
        \vspace{0pt}
        \centering
        \includegraphics[width=\textwidth]{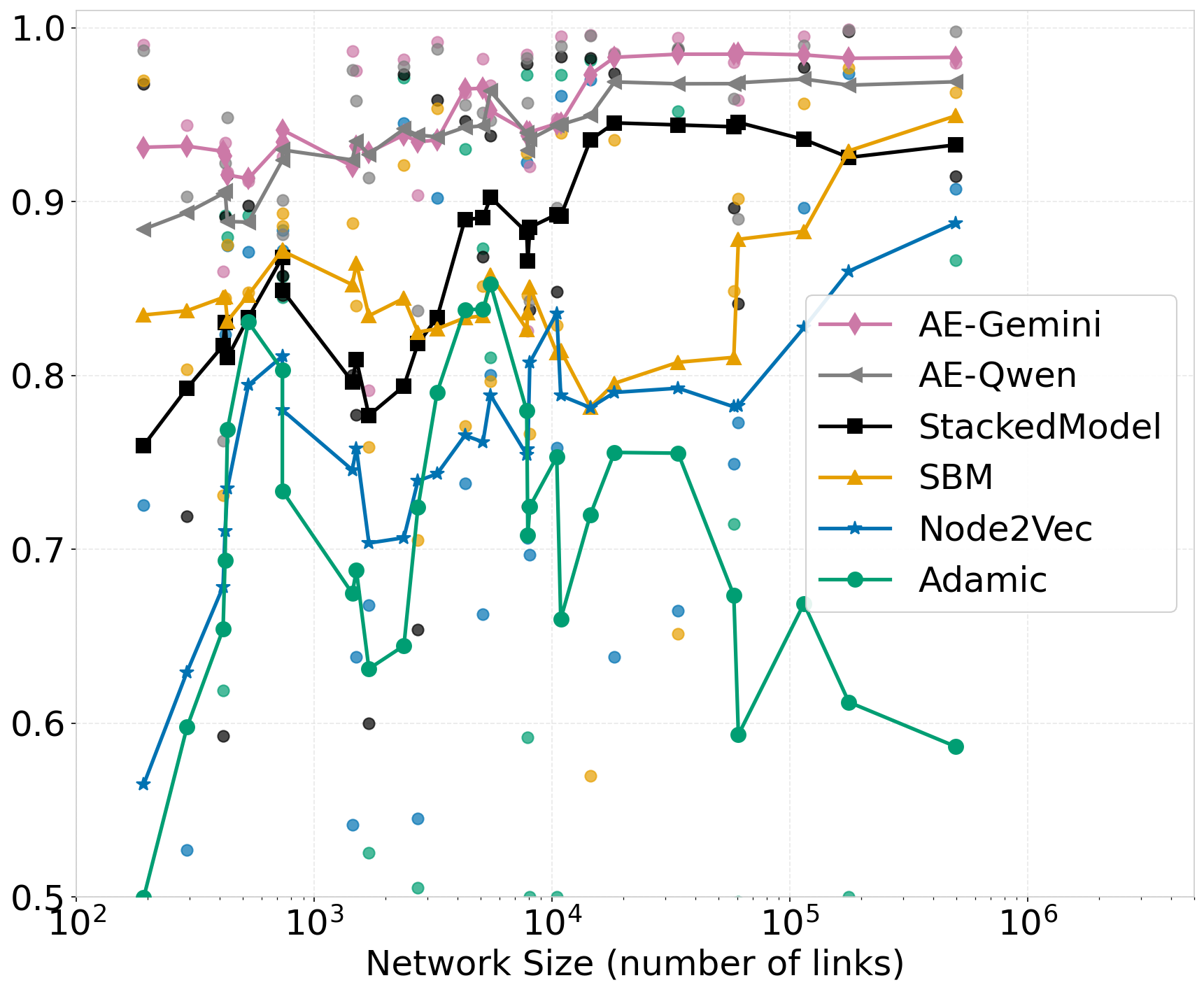}
    \end{minipage}
	\caption{\textbf{Comparison of human-designed and code evolved link-prediction methods on 30 large networks.}
	 Left panel: Overall mean AUC score averaged across 600 graph instances per method. Right panel: Detailed AUC score of methods across 30 networks of different sizes, ordered by the number of links. Each dot represents the average AUC obtained over 20 random samples of links (20\% of the links considered to be unobserved). Lines represent moving average.}
	\label{fig:graphs30}
\end{figure}

In order to evaluate our method on a larger interval of network sizes, we selected a set of 30 networks of various sizes from Ref.~\cite{tiago2020}. We sample unobserved links 20 times from each of the 30 networks and present each method with the same set of network instances where $20\%$ of links are missing. None of these networks are seen at training.  This set of networks include larger networks and thus brings to surface some of the computational challenges that algorithms face when working with larger networks.
The results summarised in Fig.~\ref{fig:graphs30} (see SM Table~\ref{tab:30nets} for all scores) reveal once more that the best performing methods found by code evolution surpasses the performance of human-designed methods. As will be shown below, this is achieved without an increase in the computational time required by the method to perform link prediction in large networks. While the results are on average decisive, the figure shows also that there are (large sparse) examples of networks in which the  methods obtained by code evolution struggle (or performs worse than other methods), indicating that further increase in performance is possible.

We conclude our performance analysis by summarising in Table~\ref{tab:scores_table} the AUC scores of all methods in the three ensembles of networks.
This provides a positive answer to our first question: code-evolution systems are indeed capable of exploring a vast landscape of algorithms for link prediction and of producing competitive algorithms for this problem.

\begin{table}[htbp]
\centering
\begin{tabular}{|c|c|c|c|}
\hline
\textbf{ \/ } & \textbf{8 Synthetic Networks } & \textbf{30 Networks} & \textbf{550 Networks} \\ \hline
Adamic & 0.490 & 0.695 & 0.641 \\
Node2Vec & 0.554 & 0.768 & 0.635 \\
SBM & 0.746 & 0.851 & 0.765 \\
Stacked model & 0.634 & 0.869 & 0.778 \\
AE-Qwen & \textbf{0.803} & 0.940 & 0.877 \\
AE-Gemini & 0.755 &  \textbf{0.951} &  \textbf{0.913} \\ \hline
\end{tabular}
\caption{\textbf{Performance comparison.} The average prediction performance (AUC score) of the six link-prediction methods in the three ensembles of test networks.}
\label{tab:scores_table}
\end{table}

\section{Discussion}

The results of the previous section confirm the success of our best AntEvolve programs $p^*$ in obtaining high AUC scores in widely different types of networks. The goal of this section is to better understand how $p^*$ and our code evolution system AntEvolve achieve this success.

\subsection{Ablation studies}

\begin{figure*}[htbp]
    \centering
    \begin{minipage}[t]{0.7\textwidth}
    \centering
        \includegraphics[width=\textwidth]{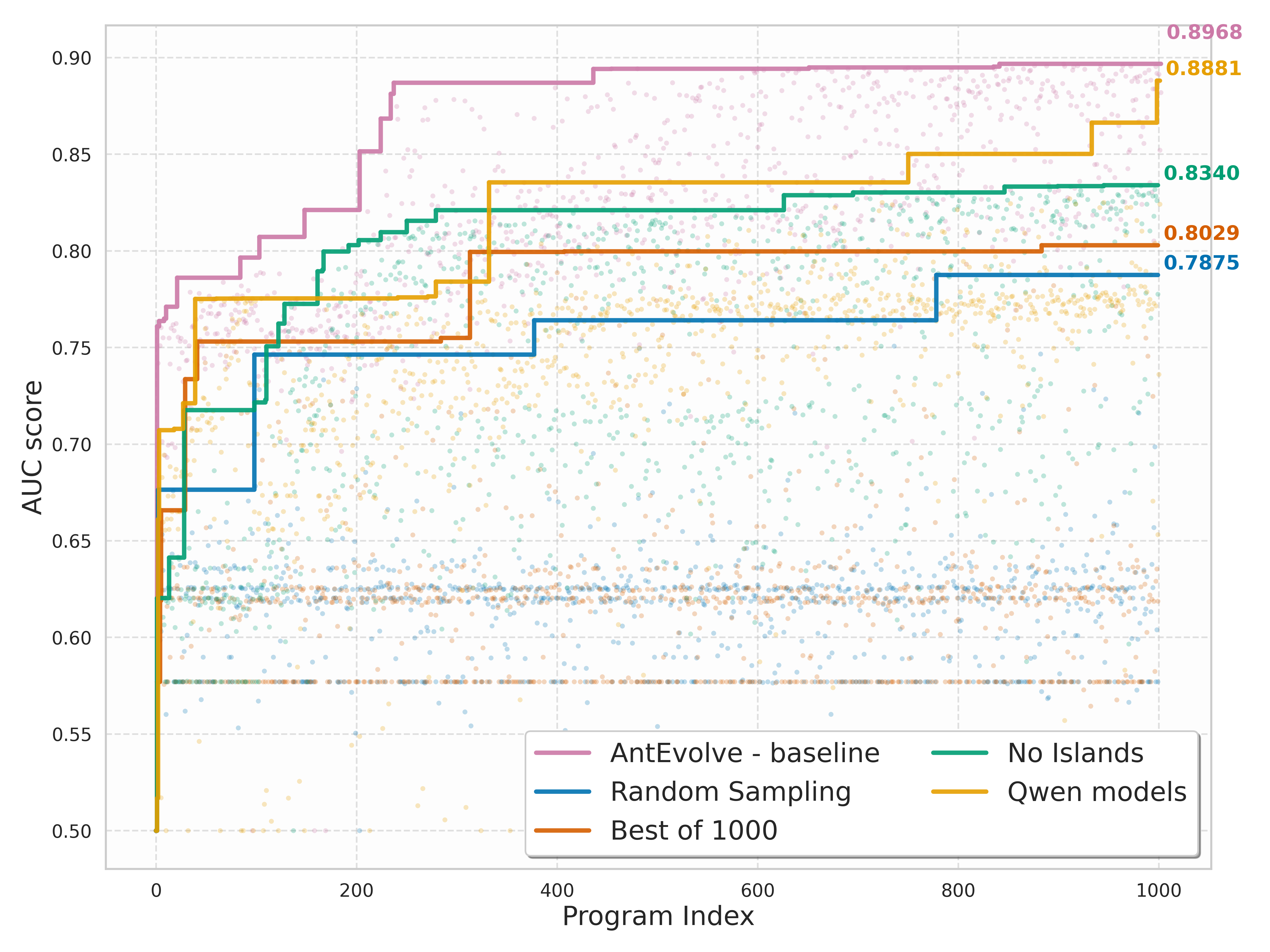}
    \end{minipage}
    \caption{\textbf{Code-evolution ablation studies.} Comparison of one realisation of the AntEvolve system with the four different configurations considered in our ablation study (see text and legend). Due to the high cost of running the system, we ran these ablation studies only once.
     }\label{fig.ablationgraph}
\end{figure*}

We conduct our ablation studies focusing on the second research question posed in the Introduction: which parts of AntEvolve's code evolution are important? We run different configurations of the code-evolution system to empirically study how its different parts and configurations affect the final performance of the best program. We consider the following  configurations:

\begin{enumerate}
    \item \textbf{Best of M.}  New programs $\bar{p}$ are proposed from mutations of the same initial program $p_0$. This corresponds to the disabling of the sequential program improvement described in Sec.~\ref{sec:codevolve}.
    \item \textbf{Random selection.} Random programs in $A_t$ are selected for mutation/crossover, creating a sequential improvement process of randomly selected programs. This corresponds to disabling the score-based selection of $S_t$ described in Sec.~\ref{sec:codevolve}.
    \item \textbf{No islands.} The number of islands is set to one in the system, which reduces the diversity of programs.
    \item \textbf{Training data.} We use four real networks by default (data A configuration), here we only use one real network with synthetic networks (data B configuration) as the code evolution evaluation data.
\end{enumerate}
We use Gemini as the default LLM choice (AntEvolve baseline) because of its highest efficiency in generating acceptable programs in comparison to Qwen.

Figure~\ref{fig.ablationgraph} shows a comparison of the evolution of programs for each of these configurations. Compared to the original AntEvolve system, all alternative configurations show a significant drop in the training score, suggesting that the elements they remove from AntEvolve do play an important role in its success.

\subsection{Robustness and efficiency}

\begin{figure*}[ht]
        \centering
    \begin{minipage}[t]{0.275\textwidth}
        \vspace{0pt}
        \centering
        \includegraphics[width=\textwidth]{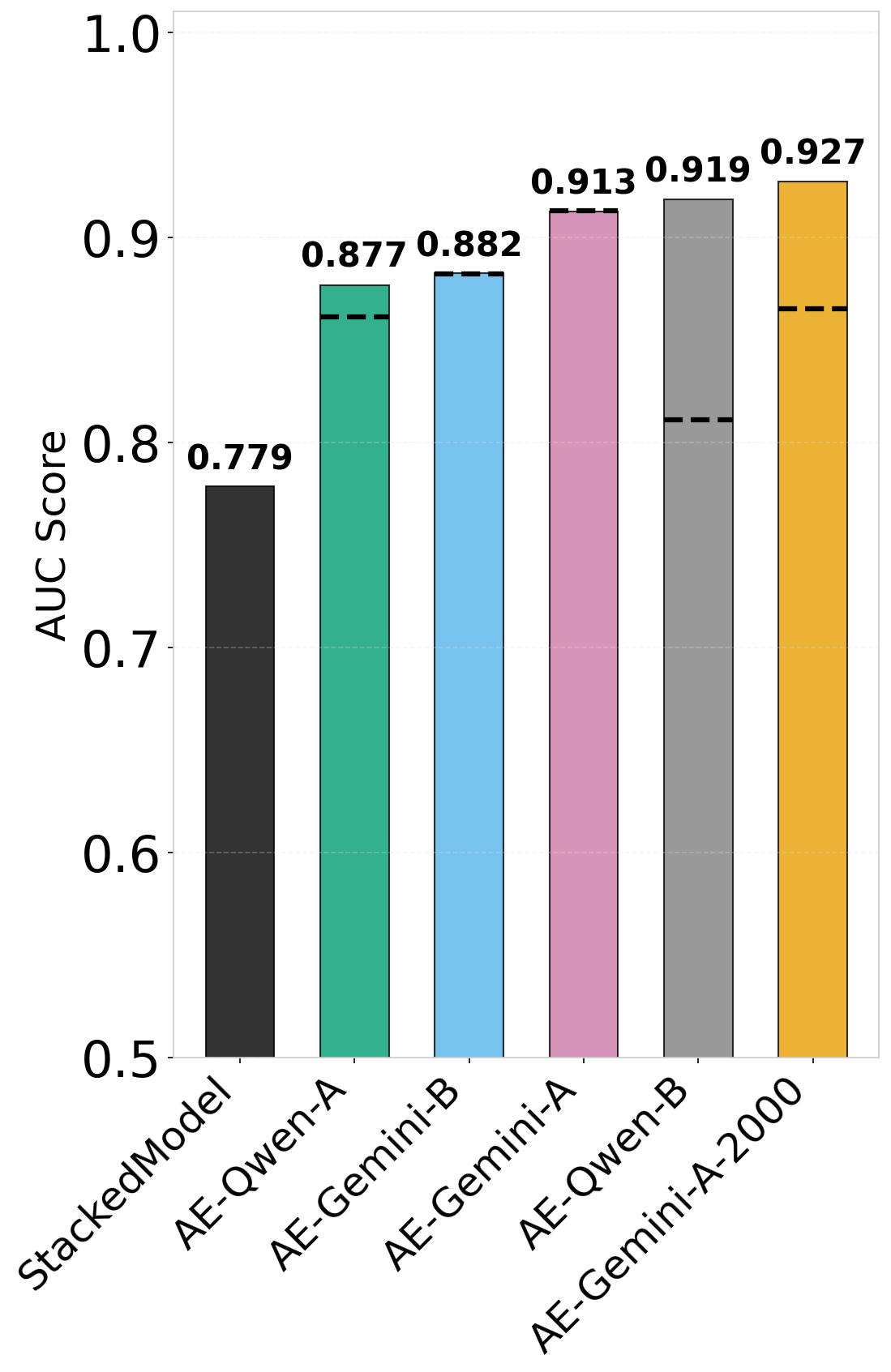}
    \end{minipage}
    \begin{minipage}[t]{0.55\textwidth}
        \vspace{0pt}
        \centering
        \includegraphics[width=\textwidth]{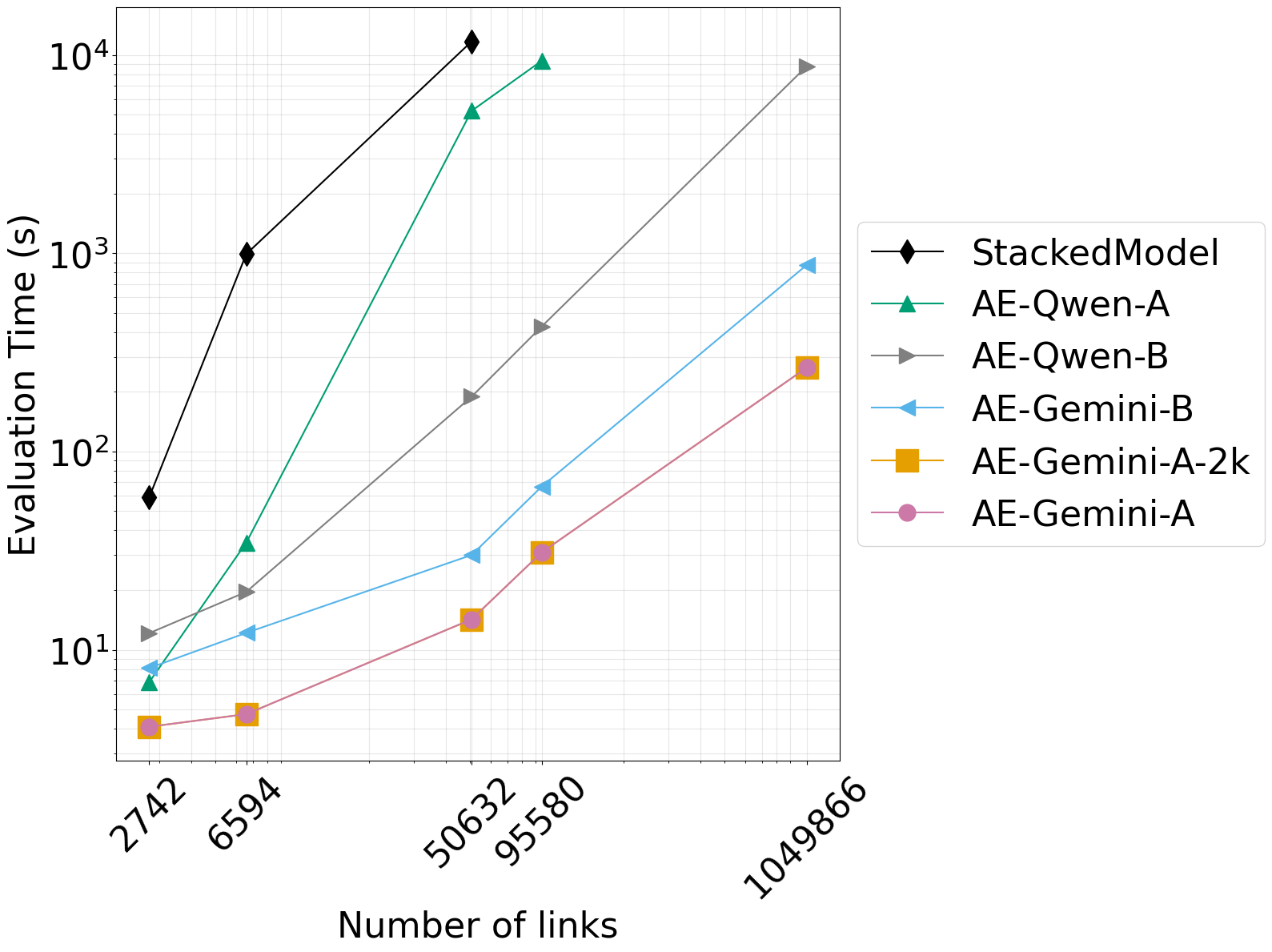}
    \end{minipage}
    \caption{\textbf{Evaluation of machine-evolved link-prediction programs.} Left panel: the average AUC score -- computed on the ensemble of $550$ test networks used in Ref.~\cite{pnas2020} and Fig.~\ref{fig.graphs550} -- of the human-designed Stacked Model and the machine-designed best programs $p^*$ obtained in five scenarios of our code-evolution. The dashed lines are the average AUC values obtained when the node IDs are randomised (as discussed in Sec.~\ref{ssec.discovered}).
    Right panel: computational time (in seconds) of evaluation of the different (best) programs as a function of network size (number of links), see SM~Sec.~\ref{ssec.speed} for details. The indices "A" and "B" at the end of the method label reflect different choices of training sets: version "A" has six synthetic networks and four real network and "B" has the same six synthetic networks and one real network.
    }\label{fig.antalgos}
\end{figure*}

So far we have focused on two runs of the code-evolution, which used a Gemini/Qwen LLMs and generated $M=1,000$ programs. We now investigate how efficient and robust our results are, in particular, with respect to the training data and $M$.  In particular, we consider evolutions with different training sets in the two LLM modes and extended the Gemini run up to $M=2,000$ accepted programs (which led to a further slight improvement in the training AUC score, see SM Fig.~\ref{fig.baseline4kapp}).
 Figure~\ref{fig.antalgos} shows the performance results of the best programs $p^*$ obtained in these cases, obtained in our largest ensemble of networks used as a test set. They confirm that a comparable AUC score is obtained in the five machine-evolved programs, superior to the best human-designed method considered here. Together, these results indicate the robustness of the results against different choices of $M$ and LLM. Importantly, the programs $p^*$ obtained using the Gemini LLM require substantially less computational time to run than the human-designed Stacked model and the program obtained using Qwen as LLM. This is a remarkable property because these are also the best-scoring programs, and their evolution did not explicitly select for algorithmic efficiency, beyond a general restriction on the maximum evaluation time of accepted programs. Code evolution might incorporate computationally expensive features to optimise performance scores. Specifically, many link-prediction methods scale poorly within large-scale networks due to the utilisation of features with square time complexity $O(|N|^2)$, where $N$ is a set of nodes in the network.
 For example, betweenness centrality is used by topological Stacked model and the best Qwen-A program.
 This feature is computed using the algorithm proposed in Ref.~\cite{brandes2001faster}, which has $O(|N| \times |E|)$ time complexity,
 where $E$ is the set of links so that, in sparse networks, it effectively behaves like $O(|N|^2)$. Our findings indicate that while certain
 evolved algorithms do retain these inefficiencies (like the $p^*$ found by Qwen models), the top-performing program obtained by
 Gemini does not and consequently demonstrates significant computational efficiency with high predictive accuracy.

\subsection{Discovered algorithms analysis}\label{ssec.discovered}

We now focus on our third research question: does the evolutionary process produce genuinely novel algorithms? Our experiments produced thousands of link-prediction methods. In order to understand them, we systematically inspected the codes of the top-performing and most-diverse programs.
We used feature analysis to assess the importance of a variety of features and concluded that all features in the best program are important, with variable contributions depending on the network (see SM Sec.~\ref{app:programs} for the list of all features of the best algorithm and further examples of features). The individual features of the best programs evolved by Gemini are computed with $O(1)$ or $O(k_v k_u)$ time complexity, where $k_i$ is the degree of node $i$, which makes these algorithms efficient in sparse networks.

The best algorithms found by our code-evolution system use a supervised learning approach with machine-learning classifiers applied to various node- and link-based features. This flexible architecture supports the inclusion of additional features during the evolution process and works effectively also for large feature spaces. Our analysis of the top 100 scoring programs shows a variety of approaches within this general architecture: single classifier on top of the topological features, voting or weighted classifiers combining low-variance and low-bias machine learning models ~\cite{kaggle}, and stacking (an ensemble learning method that combines multiple distinct machine learning models into a single, more powerful predictive model~\cite{stacking}). The best program found by Gemini uses a single randomised trees classifier~\cite{geurts2006extremely}, while the one found by Qwen uses linear regression as a stacking model for five classifiers combining low variance models (logistic regression) and low bias models (random forest and gradient boosting classifiers.)

We manually analysed the features used in the best programs. Compared to the Stacked model, which has 42 topological features, the best algorithms found by Gemini-A and Qwen-A use 24 and 40 features, respectively. Common features among all analysed algorithms include well-known link-prediction strategies (e.g. Adamic-Adar index, Jaccard coefficient, common neighbours), local information on nodes (e.g. number of triangles that a node participates in), and a variety of centrality measures (e.g. eigenvector, degree, closeness, Katz, and Pagerank). A small subset of the features used in the best algorithms appears to be novel. Among them, a few node and pair of nodes features (e.g. using combinatorial triangle-based features~\cite{shang2025triadic}, using node degree in gravity inspired feature~\cite{wei2025gaedgrn, gravity18, gravity19}) appear to be re-discoveries by the model because they have been published after January 2025 (which is the knowledge cutoff of the models we used in this research).
We also found unique features, which to our knowledge have not been proposed for link prediction (e.g. average absolute log-difference between a node's degree and the degrees of all its neighbours; weighted logarithmic node degree difference; combining node degree with number of triangles a node participates in). During the evolution process, we also noticed the proposal of nonsensical features (e.g. associating the node-labels to zodiac symbols or to digits of Pi), which were subsequently removed through the evolution process and were absent from the top-scoring programs.

A surprising discovery of our analysis is that there are predictive features based on the node identifiers (IDs) used to describe the network (e.g. the labels used in the edge list or the row/column numbers of an adjacency matrix).  For example, the best Qwen-A program uses node ID difference as a feature (see feature 38 in SM Tab.~\ref{tab:aqwen_features_two}). If node IDs were arbitrary labels, the performance of link prediction methods would remain invariant under their randomisation. Surprisingly, we find that this is not the case and that the performance of Qwen-A in our test set drops significantly under random shuffling of IDs, while the performance of the best Gemini-A program -- which does not rely on such features -- remains the same (see SM-Fig.~\ref{fig:random550}). This shows that node IDs have predictive information in many empirical networks, that AntEvolve succeeded in mapping this information into a feature, and that some of the best-found programs $p^*$ are exploiting it for successful predictions. In particular, node IDs are informative in the synthetic networks used in the training stage and in Sec.~\ref{ssec.synthetic}, explaining why AntEvolve developed such features and the success of Qwen-A in the synthetic data (compare Fig.~\ref{fig.syntheticNets} with the randomised version in SM-Fig.~\ref{fig:randomdrop}). The overall success of AntEvolve and of our reference link prediction method $p^*$ (Gemini-A) does not rely on this information (see Fig.~\ref{fig.antalgos} for the performance of all $p^*$ methods after randomising the IDs).

\section{Conclusion}

In this work, we provide empirical evidence that modern code-evolution systems are capable of exploring large program spaces and finding efficient link-prediction algorithms.
Our best machine-designed algorithms showed both better prediction scores and computational efficiency than all human-based algorithms tested. This finding was robust across different network ensembles, including 580 different empirical and synthetic networks. Remarkably, this success was achieved using only ten unique networks during the evolution of the method (training), with no selection based on computational efficiency (apart from a cut-off in the maximum computational time). The significance of this finding is that we provide both best programs $p^*$ (which have direct applications to all practical problems involving link prediction) and an open source recipe for finding them (via code-evolution). Moreover, the improved computational performance implies that the method $p^*$ can be used in much larger networks to predict links in sparse networks with millions of links.

Our ablation studies showed the importance of fitness-based selection and genetic-diversity preservation (islands) strategies used in AntEvolve - our code-evolution system. While prior work in different problems~\cite{randomllm} indicated that specific mathematical heuristics can be found without complexities of code evolution, our study suggests that surpassing state-of-the-art results in link prediction requires all components of code-evolution system. In particular, we find that islands lead to a higher diversity of programs and improved performance, in line with the theory proposed in Ref.\cite{tanese1989distributed}. Our confirmation of the importance of sequential improvement of the top scoring programs is in agreement with the results reported in AlphaEvolve~\cite{alphaevolve}.
While these are important elements in our code-evolution system, the finding of highly-efficient algorithms is robust against choices of other parameters such as random realisations of link removal, the stopping time $M$, and the choice of LLM (provided they are capable of writing Python programs).

Our findings in the link-prediction problem provide important insights into the more general question of the potential of LLM-based code-evolution systems to discover novel algorithms. A  fundamental question in this context is: to what extent code evolution can enable discovery of fully novel algorithms and what are the key enabling factors for that to happen. In our study, we observe that the general structure of the proposed algorithm  -- a supervised learning approach~\cite{al2006link} that can flexibly accommodate a variety of features - resembles a classical machine learning approach used for link-prediction. While no radically novel algorithm architecture was proposed, substantial innovation is found in the proposal and in the combination of nodal and link features.
This approach was sufficient to obtain a significant improvement over existing methods, suggesting that a recombination and improvement of human-designed algorithms was the preferred innovation path in this case.

\section*{Acknowledgements}
We thank Tristram Alexander and Yuanming Tao for providing valuable feedback on a previous version of this manuscript.

\paragraph*{Author contributions:}
A.V. and E.G.A designed the research. A.V. wrote the software for code-evolution system and performed the computations. A.V. and E.G.A. analysed the data. A.V. and E.G.A wrote the manuscript.

\paragraph*{Competing interests:}
There are no competing interests to declare.

\paragraph*{Data and materials availability:}
The source code, data, and final algorithms will be made available: all data used in this work in Ref.~\cite{datarepo}; code and results in Ref.~\cite{coderepo}. All are shared with the MIT licence.

\newpage

\renewcommand{\thefigure}{S\arabic{figure}}
\renewcommand{\thetable}{S\arabic{table}}
\renewcommand{\theequation}{S\arabic{equation}}
\renewcommand{\thepage}{S\arabic{page}}
\setcounter{figure}{0}
\setcounter{table}{0}
\setcounter{equation}{0}
\setcounter{page}{1}

\begin{center}
\setcounter{section}{0}
\section*{Supplementary Materials for\\ \mytitle}

Alexey Vlaskin$^{\ast}$,
Eduardo G. Altmann,
\\
\small$^\ast$Corresponding author. Email: alex@avlaskin.com\\
\end{center}

\subsubsection*{This PDF file includes:}
Link-prediction evaluation\\
Training networks\\
Ablation and feature analysis\\
Figures S1 to S3\\
Tables S1 to S6

\newpage

\section{Link-prediction evaluation}\label{app:auc}

In the evaluation of the AUC score, we sample an equal number of positive unobserved links and non existing links, so that we have balanced number of positive and negative examples at evaluation time. From each network in this study we randomly sample 10-20\% of existing links 10-20 times and average AUC score across those 10-20 instances.  It is critical here to keep the same set of sampled links for methods comparison because this ensures fairness of the comparison.  As Ref.~\cite{graphevalnn} pointed out, the comparison of methods should have exactly the same links removed for all methods to have the same complexity, in line with our approach. To do that, we store sampled networks with training sets and test sets of links/labels in our repository~\cite{datarepo} and use them for all methods. We disregard all the node- and edge-level features that network data come with and only use nodes and links as training and test data for link prediction.

We consider that the output of link-prediction methods is a set of predicted links $E_p \subseteq E_{U} \cup \widetilde{E}$. The size of this set, $0 \le |E_p| \le |E_{U} \cup \widetilde{E}|$, is typically controlled by a threshold $0\le t \le 1$ that sets the rate of link prediction (e.g. $E_p$ is composed by all links predicted with probability $p\ge t$ or by the top $t$ percentage of most-likely links). The number of true positives (TP) is given by $|E_p(t) \cap E_{U}|$, false positives (FP) by $|E_p(t) \cap \widetilde{E}|$, true negatives (TN) by $|\widetilde{E} \setminus E_p(t)|$, and false negatives (FN) by $|E_{U} \setminus E_p(t)|$. The true positive rate (TPR) and false positive rate (FPR) are then defined as
\begin{equation}
TPR(t) = \frac{TP(t)}{TP(t) + FN(t)},
\end{equation}
\begin{equation}
FPR(t) = \frac{FP(t)}{FP(t) + TN(t)}.
\end{equation}
In this paper we use as a measure of the quality of the link prediction the Area Under the Curve ($0\le AUC\le 1$) formed in the FPR(t) vs. TPR(t) graph when varying $t$ in $[0,1]$, defined as
\begin{equation}
AUC = \int_{0}^{1} TPR(FPR^{-1}(t)) \,dt.
\end{equation}
We use the AUC measure because it is threshold free, normalised (the larger the $AUC$, the better the prediction), and it has a well-defined value $AUC=0.5$ for a random prediction (null model).

In the evaluation of the AUC, we sample equal number of positive unobserved links and non existing links, so that we have a balanced number of positive and negative examples at evaluation time. This is critical to ensure fairness in the comparison and is aligned with the recommendation in Ref.~\cite{graphevalnn}. It is also important to mention that we construct ten graphs per each chosen configuration and show their average and variance of method performance on these graphs.

\section{Training networks}\label{app:training}

During the training phase (or code-evolution phase) we use ten unique networks for evaluation: we take six synthetic networks from Ref.~\cite{vlaskin2025synthetic} and four real networks(each instance has 20\% of links unobserved): Jazz collaboration network, Crime network, London Transportation Network, and Bible nouns network. We choose these four networks because they are small (allow for quick evaluation) and challenging for conventional methods like Node2Vec and SBM. Three synthetic networks are: a set of lattice structures with bridge nodes connecting to them, as shown on Fig.\ref{fig.syntheticNets}; another three networks use cliques as structures similarly connected by bridge nodes. Each network in the training set has three instances in which 20\% of links are removed for evaluation.

\begin{table}[htbp]
\centering
\footnotesize
\begin{tabular}{llrr}
\hline
\textbf{Network} & \textbf{Nodes} & \textbf{Edges} & \textbf{Average Degree} \\
\hline
revolution & 141 & 160 & 2.27 \\
interactome\_pdz & 212 & 244 & 2.3  \\
marvel\_partnerships & 350 & 346 & 1.98 \\
game\_thrones & 107 & 352 & 6.58  \\
student\_cooperation & 185 & 360 & 3.89  \\
polbooks & 105 & 441 & 8.4 \\
football & 115 & 613 & 10.66 \\
football\_tsevans & 115 & 613 & 10.66 \\
plant\_pol\_kato & 772 & 1,206 & 3.12 \\
unicodelang & 868 & 1,255 & 2.89 \\
euroroad & 1,174 & 1,417 & 2.41 \\
facebook\_friends & 362 & 1,988 & 10.98 \\
interactome\_yeast & 1,870 & 2,277 & 2.44 \\
netscience & 1,589 & 2,742 & 3.45 \\
eu\_airlines & 450 & 3,588 & 15.95 \\
new\_zealand\_collab & 1,511 & 4,273 & 5.66 \\
celegans\_metabolic & 453 & 4,596 & 20.29\\
wiki\_science & 687 & 6,523 & 18.99 \\
power & 4,941 & 6,594 & 2.67 \\
interactome\_vidal & 3,133 & 6,726 & 4.29  \\
drosophila\_flybi & 2,939 & 8,723 & 5.94\\
collins\_yeast & 1,622 & 9,070 & 11.18 \\
dnc & 2,029 & 12,085 & 11.91 \\
plant\_pol\_robertson & 1,884 & 15,265 & 16.20 \\
contact & 274 & 28,244 & 206 \\
internet\_as & 22,963 & 48,436 & 4.22 \\
escorts & 16,730 & 50,632 & 6.05 \\
marvel\_universe & 19,428 & 95,497 & 9.83 \\
reactome & 6,327 & 147,547 & 46.64 \\
sp\_infectious & 10,972 & 415,912 & 75.81 \\
dblp\_com & 425,957 & 1,049,866 & 4.93 \\
\hline
\end{tabular}
\caption{Ensemble of 30 networks used to test link prediction methods. The dblp\_com network is used only for efficiency tests.}\label{tab:stat30nets}
\end{table}

\begin{table}[htbp]
\centering
\footnotesize
\begin{tabular}{lrrrrrr}
\hline
\textbf{Network} & \textbf{Adamic} & \textbf{Node2Vec} & \textbf{SBM} & \textbf{Stacked model} & \textbf{AntEvolve} & \textbf{AE-Qwen} \\
 & \textit{AUC(std)} & \textit{AUC(std)} & \textit{AUC(std)} & \textit{AUC(std)} & \textit{AUC(std)} & \textit{AUC(std)} \\
\hline
revolution & 0.39(05) & 0.73(12) & 0.97(03) & 0.97(04) & 0.99(02) & 0.98(02) \\
interactome\_pdz & 0.49(02) & 0.53(10) & 0.80(06) & 0.72(11) & 0.94(03) & 0.90(05) \\
marvel\_partnerships & 0.62(03) & 0.44(08) & 0.73(06) & 0.59(07) & 0.86(04) & 0.81(06) \\
game\_thrones & 0.89(03) & 0.82(06) & 0.84(04) & 0.89(04) & 0.93(02) & 0.88(03) \\
student\_cooperation & 0.88(04) & 0.87(04) & 0.88(04) & 0.92(04) & 0.92(04) & 0.94(03) \\
polbooks & 0.89(03) & 0.87(03) & 0.85(04) & 0.90(03) & 0.91(03) & 0.94(02) \\
football & 0.85(03) & 0.87(03) & 0.89(03) & 0.85(04) & 0.93(02) & 0.91(03) \\
football\_tsevans & 0.86(04) & 0.88(03) & 0.89(04) & 0.86(04) & 0.93(02) & 0.90(03) \\
plant\_pol\_kato & 0.45(01) & 0.54(06) & 0.89(03) & 0.80(13) & 0.99(01) & 0.98(01) \\
unicodelang & 0.48(01) & 0.64(06) & 0.84(03) & 0.78(09) & 0.98(01) & 0.97(01) \\
euroroad & 0.53(01) & 0.67(05) & 0.76(03) & 0.60(06) & 0.79(03) & 0.93(02) \\
facebook\_friends & 0.97(01) & 0.95(01) & 0.92(02) & 0.97(01) & 0.98(01) & 0.98(01) \\
interactome\_yeast & 0.51(02) & 0.55(05) & 0.71(03) & 0.65(05) & 0.90(02) & 0.93(01) \\
netscience & 0.94(01) & 0.90(02) & 0.95(01) & 0.96(01) & 0.99(01) & 0.99(01) \\
eu\_airlines & 0.93(01) & 0.74(03) & 0.77(01) & 0.95(01) & 0.96(01) & 0.97(01) \\
new\_zealand\_collab & 0.87(01) & 0.66(03) & 0.85(02) & 0.87(05) & 0.98(01) & 0.98(01) \\
celegans\_metabolic & 0.81(18) & 0.80(02) & 0.80(02) & 0.94(01) & 0.97(01) & 0.96(01) \\
wiki\_science & 0.97(01) & 0.92(01) & 0.93(01) & 0.98(01) & 0.98(01) & 0.98(01) \\
power & 0.59(01) & 0.71(02) & 0.85(01) & 0.72(05) & 0.83(01) & 0.98(01) \\
interactome\_vidal & 0.50(01) & 0.70(02) & 0.77(01) & 0.84(02) & 0.92(01) & 0.93(01) \\
drosophila\_flybi & 0.50(01) & 0.76(02) & 0.83(01) & 0.85(01) & 0.95(01) & 0.93(01) \\
collins\_yeast & 0.97(01) & 0.96(01) & 0.94(01) & 0.98(01) & 1.00(01) & 0.99(01) \\
dnc & 0.98(01) & 0.97(01) & 0.57(02) & 0.98(01) & 1.00(01) & 0.99(01) \\
plant\_pol\_robertson & 0.41(01) & 0.64(02) & 0.94(01) & 0.97(01) & 0.99(01) & 0.98(01) \\
contact & 0.95(01) & 0.66(03) & 0.65(04) & 0.99(01) & 0.99(01) & 0.99(01) \\
internet\_as & 0.71(01) & 0.75(02) & 0.85(01) & 0.90(05) & 0.98(01) & 0.99(01) \\
escorts & 0.50(01) & 0.77(02) & 0.90(01) & 0.84(03) & 0.96(01) & 0.93(01) \\
marvel\_universe & 0.48(01) & 0.90(01) & 0.96(01) & 0.98(01) & 1.00(01) & 0.98(01) \\
reactome & 0.50(01) & 0.97(01) & 0.98(01) & 1.00(01) & 1.00(01) & 1.00(01) \\
sp\_infectious & 0.87(01) & 0.91(01) & 0.96(01) & 0.91(01) & 0.98(01) & 1.00(01) \\
\hline
\end{tabular}
\caption{AUC scores of six link prediction methods across 30 networks. Values shown as mean(std).}
\label{tab:30nets}
\end{table}

\section{Ablation and feature analysis}

\subsection{Speed assessment}\label{ssec.speed}
In order to quantify the speed of program evaluation, we selected five progressively large networks from the Netzschleuder repository~\cite{tiago2020}: NetScience with ($2,742$ links); power network ($6,594$ links);
Brazilian escort network ($50,185$ links); movielens 100k network ($95,580$ links); and DBLP co-authorship network ($1,049,866$ links).
We ran all assessment on the same computer instance (with 4 CPUs and 32GB of RAM) ten times. For the Stacked model, 128GB of RAM were needed in the case of the network (movielens\_100k). We then average the computation time over the 10 runs and plotted it in Figure \ref{fig.antalgos}.

\begin{figure*}[!hbtp]
        \centering
    \begin{minipage}[t]{0.8\textwidth}
        \centering
        \includegraphics[width=\textwidth]{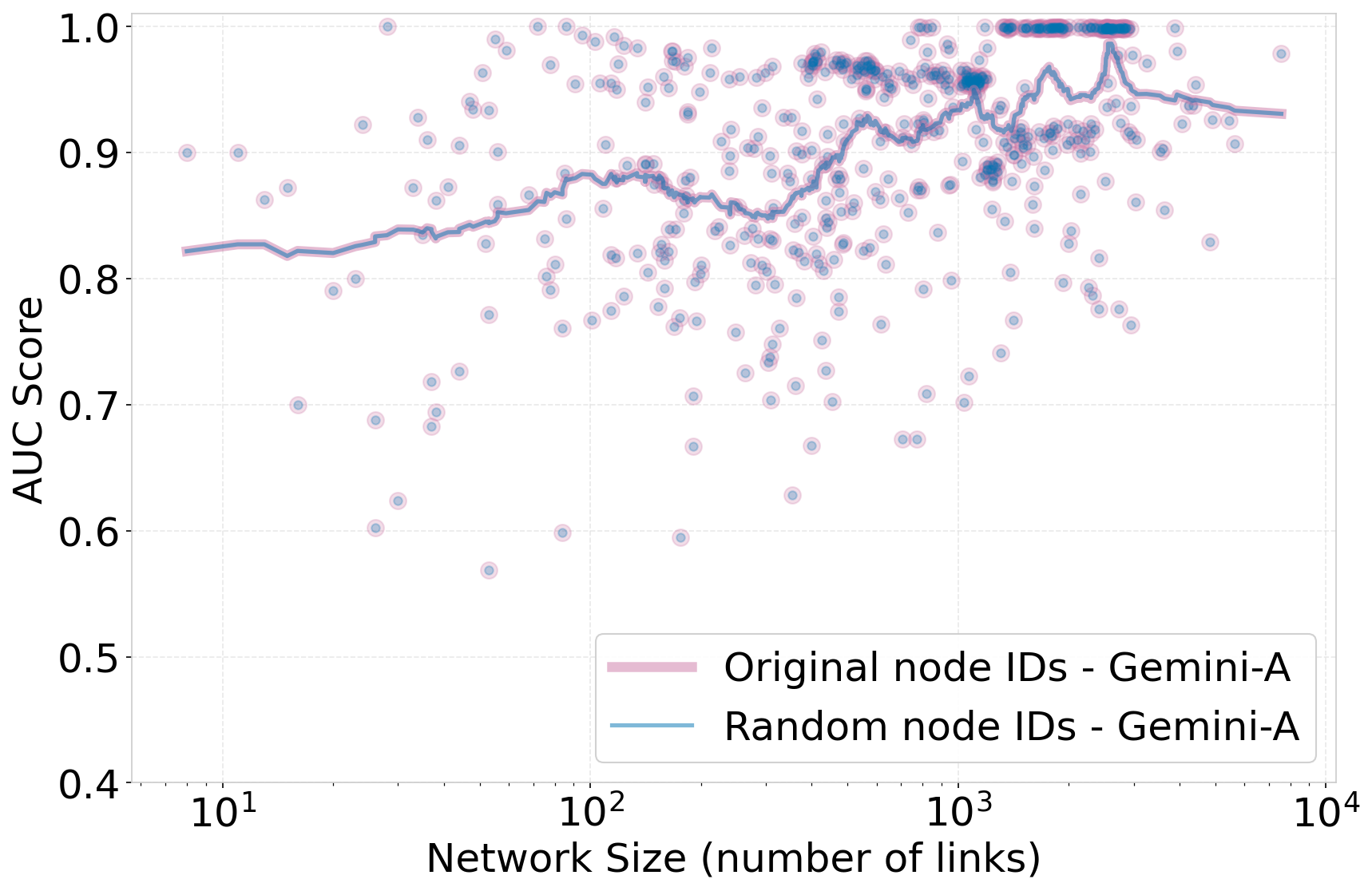}
    \end{minipage}
    \begin{minipage}[t]{0.8\textwidth}
        \centering
        \includegraphics[width=\textwidth]{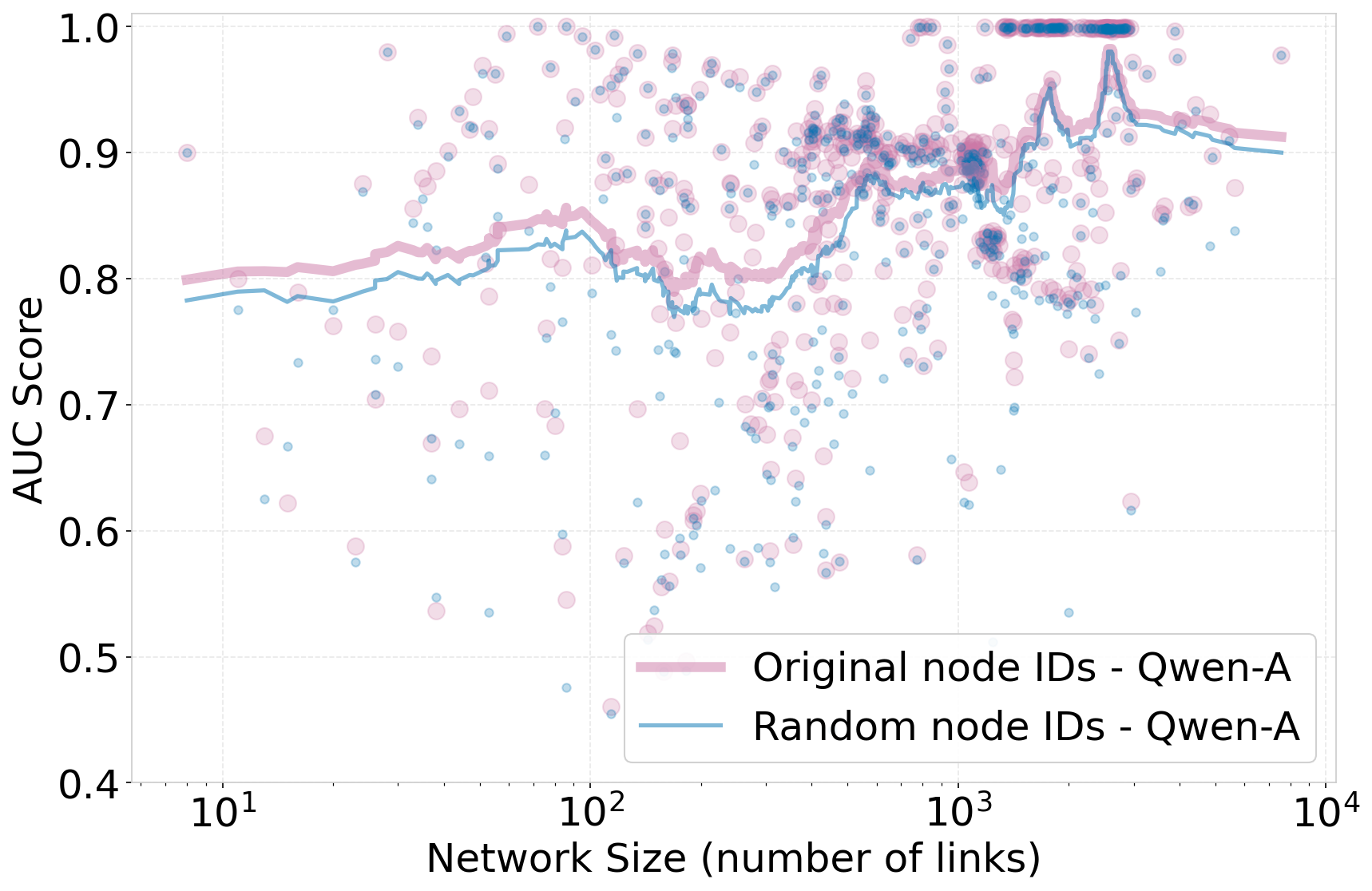}
    \end{minipage}
    \caption{\textbf{Evaluation of machine-evolved link-prediction programs with node ID randomisation.} Before evaluating the method, we randomly shuffle the node IDs of each of the test networks. The data used in this figure correspond to the 550 networks discussed in Sec.~\ref{ssec.550}. Top panel: Gemini-A method original score compared to one evaluated after random shuffling node IDs. No score changes are visible. Bottom panel: Qwen-A original score compared to the one evaluated after random node ID mapping. A noticeable drop in the score is observed.
    }\label{fig:random550}
\end{figure*}

\begin{figure*}[!hbtp]
    \centering
    \includegraphics[width=0.8\textwidth]{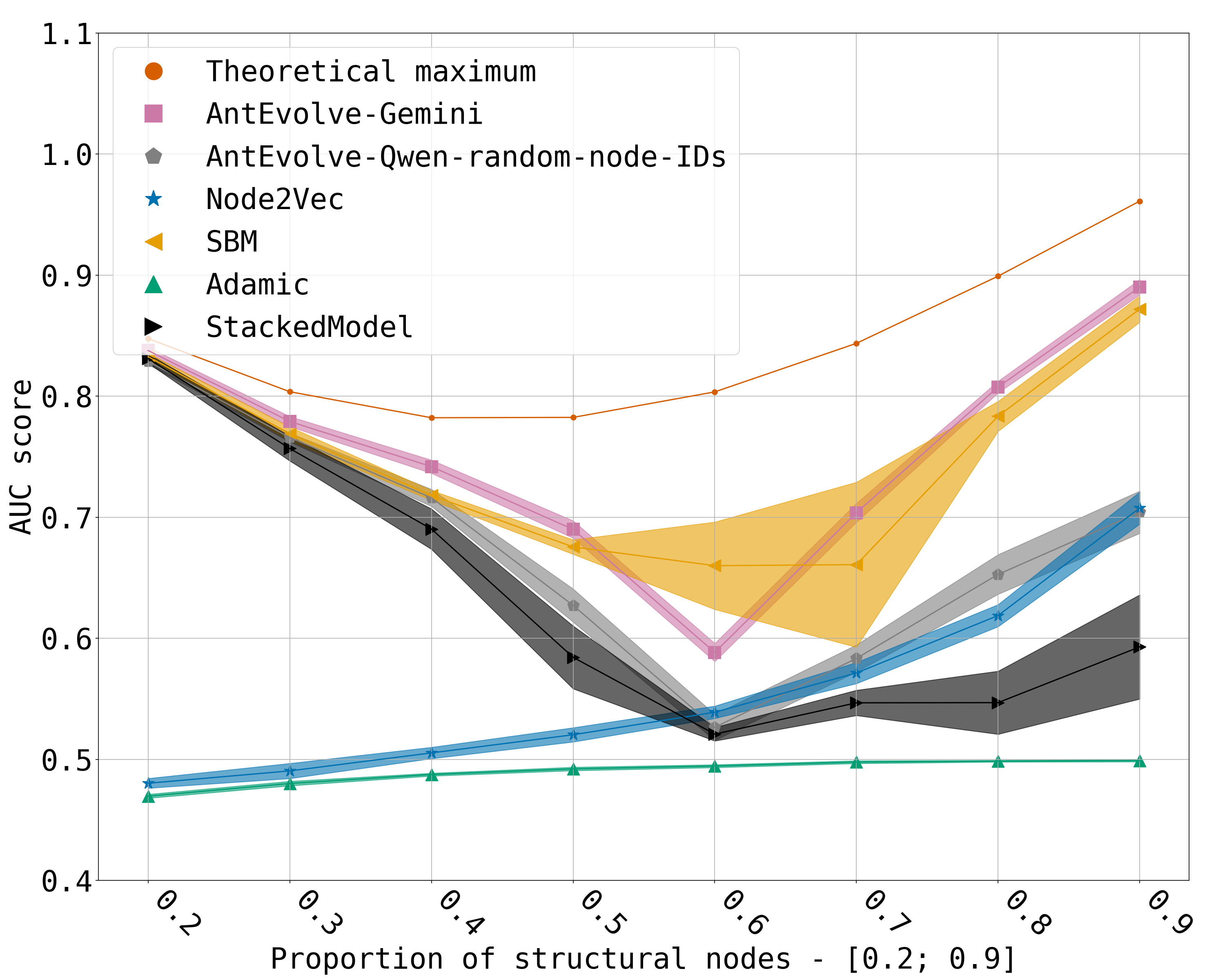}
    \caption{\textbf{Comparison of human-designed and evolved link-prediction methods in synthetic networks with node ID randomisation.} Before evaluating the method we randomly shuffle the node IDs of each of the test networks. The data used in this figure correspond to the synthetic networks discussed in Sec.~\ref{ssec.synthetic}. A drop in AUC score for Qwen-A method is observed due to lost node ID information (grey line).}
    \label{fig:randomdrop}
\end{figure*}

\subsection{Program feature analysis}\label{app:programs}

\begin{table}[htb!]
    \centering
    \footnotesize
    \begin{tabular}{|p{0.25\textwidth}|p{0.4\textwidth}|c|}
        \hline
        \textbf{Index \& Name} & \textbf{Description} & \textbf{Formula} \\
        \hline

        1. Common neighbours & Number of shared neighbours between $u$ and $v$. & $|\Gamma(u) \cap \Gamma(v)|$ \\
        2. Jaccard Coefficient & Ratio of common neighbours to the total number of unique neighbours. & $\frac{|\Gamma(u) \cap \Gamma(v)|}{|\Gamma(u) \cup \Gamma(v)|}$ \\
        3. Preferential Attachment & Natural logarithm of the product of the effective degrees plus one. & $\ln(1 + k_u \cdot k_v)$ \\
        4. Adamic-Adar Index & Sum of the inverse logarithmic degrees of all common neighbours. & $\sum_{w \in \Gamma(u) \cap \Gamma(v)} \frac{1}{\ln(k_w)}$ \\
        5. Resource Allocation & Sum of the inverse degrees of all common neighbours. & $\sum_{w \in \Gamma(u) \cap \Gamma(v)} \frac{1}{k_w}$ \\
        6. Sørensen Index & Ratio of twice the common neighbours to the sum of the nodes' degrees. & $\frac{2|\Gamma(u) \cap \Gamma(v)|}{k_u + k_v}$ \\
        7. Hub Promoted Index & Quantifies topological overlap, mitigating the dominance of high-degree nodes. & $\frac{|\Gamma(u) \cap \Gamma(v)|}{\min(k_u, k_v)}$ \\
        8. Hub Depressed Index & Similar to HPI, but penalises based on the maximum degree between $u$ and $v$. & $\frac{|\Gamma(u) \cap \Gamma(v)|}{\max(k_u, k_v)}$ \\
        9. Leicht-Holme-Newman & Ratio of actual common neighbours to the expected number in a configuration model. & $\frac{|\Gamma(u) \cap \Gamma(v)|}{k_u \cdot k_v}$ \\
        \hline
        10. Local Clustering of $u$ & Density of connections among the neighbours of $u$, based on triangle count $t_u$. & $CC_u = \frac{2t_u}{k_u(k_u - 1)}$ \\
        11. Local Clustering of $v$ & Density of connections among the neighbours of $v$, based on triangle count $t_v$. & $CC_v = \frac{2t_v}{k_v(k_v - 1)}$ \\
        12. Product of local clustering & Interaction term representing the joint local clustering probability. & $CC_u \cdot CC_v$ \\
        13. Average local clustering & The arithmetic mean of the local clustering coefficients of $u$ and $v$. & $\frac{CC_u + CC_v}{2}$ \\

        \hline
    \end{tabular}
    \caption{Features between node u and node v that are found in $p^*$ evolved by Gemini with dataset A.}
    \label{tab:common_features}
\end{table}

\begin{table}[htbp]
    \centering
    \footnotesize
    \begin{tabular}{|p{0.25\textwidth}|p{0.3\textwidth}|c|}
        \hline
        \textbf{Index \& Name} & \textbf{Description} & \textbf{Formula} \\
        \hline

        14. Weighted length-3 path & Weighted sum of length-3 paths, penalised by intermediate node degrees. & $\sum_{\substack{w \in \Gamma(u), z \in \Gamma(v) \\ w \sim z}} \frac{1}{\ln(1+k_w)\ln(1+k_z)}$ \\
        15. Path-3 Count & Total number of distinct length-3 paths connecting $u$ and $v$. & $|\{ (w, z) \mid w \in \Gamma(u), z \in \Gamma(v), w \sim z \}|$ \\
        \hline
        16. Avg neighbour Degree $u$& The arithmetic mean of the degrees of $u$'s direct neighbours. & $AND_u = \frac{1}{k_u} \sum_{w \in \Gamma(u)} k_w$ \\
        17. Avg neighbour Degree $v$& The arithmetic mean of the degrees of $v$'s direct neighbours. & $AND_v = \frac{1}{k_v} \sum_{w \in \Gamma(v)} k_w$ \\
        18. Min AND & The minimum of the average neighbour degrees of $u$ and $v$. & $\min(AND_u, AND_v)$ \\
        19. Max AND & The maximum of the average neighbour degrees of $u$ and $v$. & $\max(AND_u, AND_v)$ \\

        20. Min Degree & The minimum effective degree between $u$ and $v$. & $\min(k_u, k_v)$ \\
        21. Max Degree & The maximum effective degree between $u$ and $v$. & $\max(k_u, k_v)$ \\
        22. Degree Ratio & Ratio of $k_u$ to $k_v$, with $\epsilon$ added for numerical stability. & $\frac{k_u}{k_v + \epsilon}$ \\
        23. Triangle Count $u$ ($t_u$) & The total adjusted number of triangles in the network that include $u$. & $t_u$ \\
        24. Triangle Count $v$ ($t_v$) & The total adjusted number of triangles in the network that include $v$. & $t_v$ \\
        \hline
    \end{tabular}

      \caption{Features between node u and node v that are found in $p^*$ evolved by Gemini with dataset A.}
      \label{tab:agem_features}
\end{table}

 To formally describe these features let's introduce the following notation. Let $G = (N, E)$ be the network, with $N$ being set of all nodes, $E$ set of all links and let $u$ and $v$ denote the target pair of nodes being evaluated for a potential link. We define $\Gamma(x)$ as the \textit{effective neighbourhood set} of a node $x$, which corresponds to its adjacent nodes. Consequently, $k_x = |\Gamma(x)|$ represents the effective degree of node $x$. Nodes within these neighbourhoods are denoted by $w$ and $z$, with the relation $w \sim z$ indicating an existing link between them. The variable $t_x$ represents the adjusted local triangle count for node $x$ (accounting for the hypothetical removal of the $(u,v)$ edge). Furthermore, $CC_x$ and $AND_x$ refer to the modified local clustering coefficient and the average neighbour degree of node $x$, respectively. In ~\ref{tab:agem_features} and ~\ref{tab:common_features} you can see features used by the best algorithm $p^*$.

 \begin{table}[htbp]
\centering
\footnotesize
\renewcommand{\arraystretch}{1.2}
\begin{tabular}{|p{0.25\textwidth}|p{0.4\textwidth}|c|}
\hline
\textbf{Index and Name} & \textbf{Feature description} & \textbf{Formula} \\
\hline
1. Adamic-Adar (AA) & Sum of inverse logarithmic degrees of common neighbours. & $\sum_{w \in \Gamma(u) \cap \Gamma(v)} \frac{1}{\log(k_w)}$ \\ \hline
2. Jaccard Coefficient & Ratio of common neighbours to total unique neighbours. & $\frac{|\Gamma(u) \cap \Gamma(v)|}{|\Gamma(u) \cup \Gamma(v)|}$ \\ \hline
3. Resource Allocation (RA) & Sum of inverse degrees of common neighbours. & $\sum_{w \in \Gamma(u) \cap \Gamma(v)} \frac{1}{k_w}$ \\ \hline
4. Common neighbours (CN) & Number of shared neighbours between $u$ and $v$. & $|\Gamma(u) \cap \Gamma(v)|$ \\ \hline
5. Preferential Attachment & Product of the effective degrees of $u$ and $v$. & $k_u \cdot k_v$ \\ \hline
6. Salton Index & \raggedright Neighbourhood similarity between $u$ and $v$. & $\frac{|\Gamma(u) \cap \Gamma(v)|}{\sqrt{k_u \cdot k_v}}$ \\ \hline
7. SVD Embedding Similarity & \raggedright Cosine similarity of the 16-dimensional SVD node embeddings ($\vec{e}_x$). & $\frac{\vec{e}_u \cdot \vec{e}_v}{\|\vec{e}_u\| \|\vec{e}_v\|}$ \\ \hline
8, 9. Clustering Coefficient & Modified local clustering coefficients for $u$ and $v$. & $CC_u, \ CC_v$ \\ \hline
10, 11. Node Degrees & Effective degrees of $u$ and $v$. & $k_u, \ k_v$ \\ \hline
12, 13. PageRank & PageRank centrality scores ($\alpha=0.85$) for $u$ and $v$. & $PR_u, \ PR_v$ \\
14, 15. Betweenness Centrality & Betweenness centrality metrics for $u$ and $v$. & $BC_u, \ BC_v$ \\ \hline
16, 17. Closeness Centrality & Closeness centrality metrics for $u$ and $v$. & $C_u, \ C_v$ \\ \hline
18. Shortest Path Length & Minimum number of edges connecting $u$ to $v$. & $d(u, v)$ \\ \hline
19, 20. Triangle Count & Adjusted local triangle counts for $u$ and $v$. & $t_u, \ t_v$ \\ \hline
\end{tabular}
\caption{Features between node u and node v that are found in $p^*$ evolved by Qwen with dataset A.}
\label{tab:aqwen_features_one}
\end{table}

\begin{table}[htbp]
\centering
\footnotesize
\renewcommand{\arraystretch}{1.1}
\begin{tabular}{|p{0.25\textwidth}|p{0.4\textwidth}|c|}
\hline
\textbf{Index and Name} & \textbf{Feature description} & \textbf{Formula} \\
\hline
21. Katz Index (Path 2) & Number of paths of length 2 between $u$ and $v$. & $(A^2)_{uv}$ \\ \hline
22. Katz Index (Path 3) & Number of paths of length 3 between $u$ and $v$. & $(A^3)_{uv}$ \\ \hline
23. Katz Index (Path 4) & Number of paths of length 4 between $u$ and $v$. & $(A^4)_{uv}$ \\ \hline
24. Total neighbours & Size of the union of effective neighbourhoods for $u$ and $v$. & $|\Gamma(u) \cup \Gamma(v)|$ \\ \hline
25. Shared neighbour Conns. & Second-order proximity: count of existing links between shared neighbours. & $\sum_{\substack{w, z \in \Gamma(u) \cap \Gamma(v) \\ w \neq z}} \mathbb{I}(w \sim z)$ \\ \hline
26. Hub Promoted Index & Normalises CN by the minimum degree of $u$ and $v$. & $\frac{|\Gamma(u) \cap \Gamma(v)|}{\min(k_u, k_v)}$ \\ \hline
27. Hub Depressed Index & Normalises CN by the maximum degree of $u$ and $v$. & $\frac{|\Gamma(u) \cap \Gamma(v)|}{\max(k_u, k_v)}$ \\ \hline
28. Leicht-Holme-Newman & Normalises CN by the product of degrees. & $\frac{|\Gamma(u) \cap \Gamma(v)|}{k_u \cdot k_v}$ \\ \hline
29. Sørensen Index & Normalises CN by the sum of degrees of $u$ and $v$. & $\frac{2|\Gamma(u) \cap \Gamma(v)|}{k_u + k_v}$ \\ \hline
30, 31. Eigenvector Centrality & Eigenvector centrality scores for $u$ and $v$. & $EC_u, \ EC_v$ \\ \hline
32, 33. Degree Centrality & Normalised degree centrality scores for $u$ and $v$. & $DC_u, \ DC_v$ \\ \hline
34, 35. Avg. neighbour Degree & The average neighbour degree of $u$ and $v$. & $AND_u, \ AND_v$ \\ \hline
36, 37. Var. neighbour Degree & The variance of the degrees of the immediate neighbours of $u$ and $v$. & $\text{Var}_{w \in \Gamma(u)}(k_w), \ \text{Var}_{z \in \Gamma(v)}(k_z)$ \\ \hline
38. Node ID Difference & Absolute numerical difference between node IDs. & $|u - v|$ \\ \hline
39. Node ID Ratio & Ratio of the smaller node ID to the larger node ID. & $\frac{\min(u, v)}{\max(u, v)}$ \\ \hline
40. Avg. Clustering (Common) & Mean local clustering coefficient of all shared neighbours. & $\frac{1}{|\Gamma(u) \cap \Gamma(v)|} \sum\limits_{w \in \Gamma(u) \cap \Gamma(v)} CC_w$ \\ \hline
\end{tabular}
\caption{Features between node u and node v that are found in $p^*$ evolved by Qwen with dataset A.}
\label{tab:aqwen_features_two}
\end{table}

\begin{figure*}[!ht]
    \centering
        \includegraphics[width=0.9\textwidth]{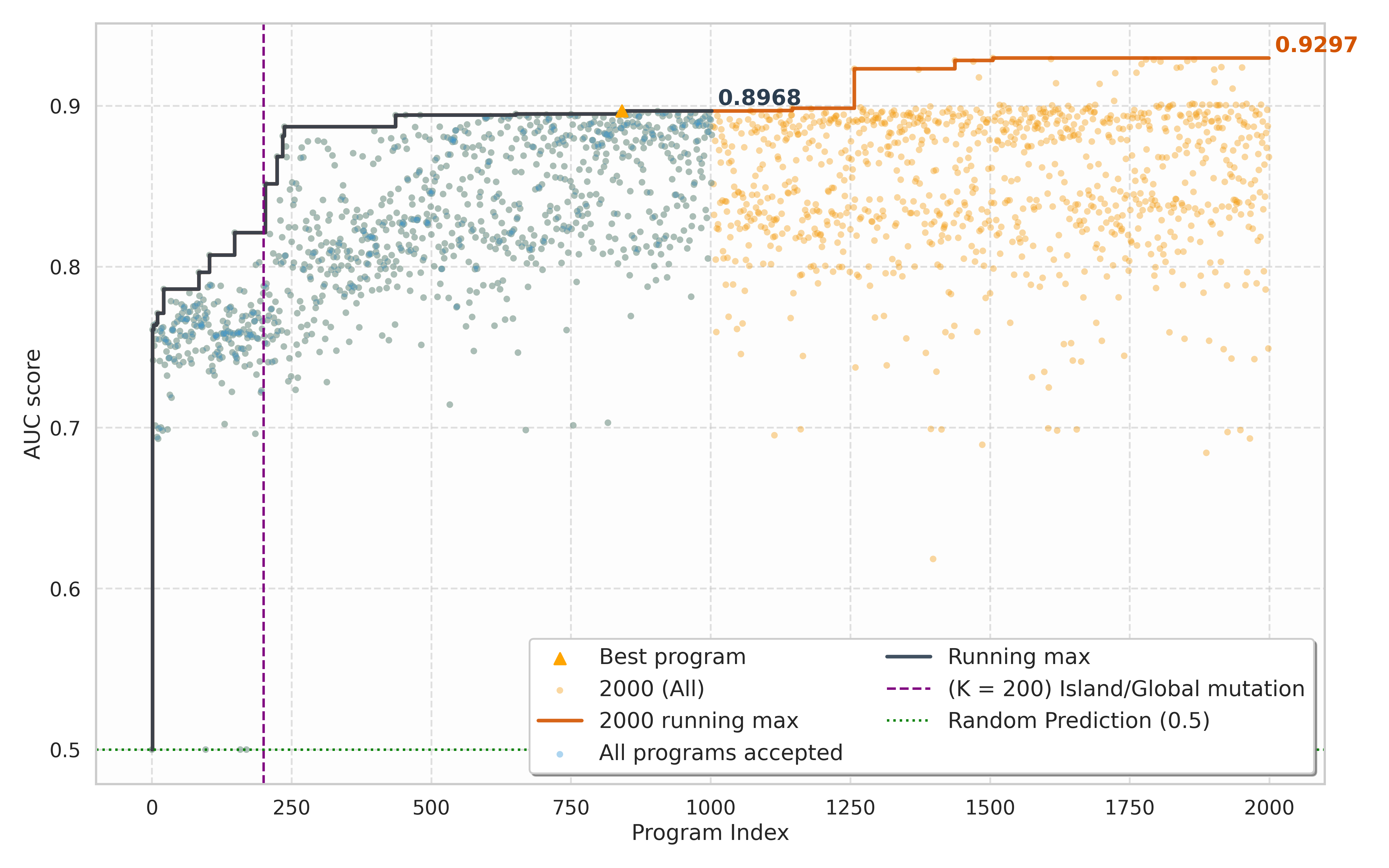}

    \caption{\textbf{Code evolution ablation study.} Expansion of the Gemini-based code evolution shown in the main text to $M=2000$ accepted programs. We see that additional training score is gained.}\label{fig.baseline4kapp}
\end{figure*}

\newpage

\end{document}